\documentclass[12pt]{iopart}
\usepackage{graphicx}
\begin{document}

\title{The inverse fallacy and quantum formalism}

\author{Riccardo Franco
\footnote[3]{To whom correspondence should be addressed
riccardo.franco@polito.it}}
\address{Dipartimento di Fisica and U.d.R. I.N.F.M., Politecnico di Torino
C.so Duca degli Abruzzi 24, I-10129 Torino, Italia}

\date{\today}

\begin{abstract}
In the present article we consider the inverse fallacy, a well known cognitive heuristic experimentally tested in
cognitive science, which occurs for intuitive judgments in situations of bounded rationality. We show that the quantum
formalism can be used to describe in a very simple and general way this fallacy within the quantum formalism. Thus we suggest that in cognitive science the formalism of quantum mechanics can be used to
describe a \textit{quantum regime}, the bounded-rationality regime, where the cognitive heuristics are valid.
\end{abstract}

\maketitle
\section{Introduction}
This paper considers how the inverse fallacy can have a natural explanation 
by using the quantum formalism to describe estimed probabilities.
We have written this article in order to be readable both from 
quantum physicists and from experts of cognitive science.
Quantum mechanics, for its counterintuitive
predictions, seems to provide a good formalism to describe puzzling effects such as 
the cognitive heuristics. In particular, violations of Bayes' rule have been predicted form quantum formalism 
\cite{Rrfranco2}.

More generally, a number of attempts have been done to apply the formalism of
quantum mechanics to domains of science different from the micro-world with applications to
economics \cite{economy}, operations research and management sciences \cite{management} and \cite{Rfranco_rat_ign_1},
psychology and cognition \cite{psychology} and \cite{aerts3}, game theory \cite{game}, and language and
artificial intelligence \cite{language}.

This paper is organized as follows:
first we recall the Bayes' theorem and the main results about
inverse fallacy and base-rate fallacy. Then we define the main
concepts of quantum formalism and show how the inverse fallacy
founds a natural explanation. Finally, we note that the conjunction fallacy
and the inverse fallacy have,  by means of the quantum formalism,
a common origin.

\section{The Bayes' theorem and the inverse fallacy}
%
The Bayes' theorem is an important result in probability theory which 
tells how to update or revise beliefs in light of new evidence a posteriori.
Suppose for example that a clinician  knows that \\
(a) only 5\% of the overall population suffers from disease X, \\
(b) 85\% of patients who have the disease show the symptom, and \\
(c) 25\% of healthy patients show the symptom. \\
We can rewrite these data in terms of probabilities:
$P(D)=0.05$ the probability that the patient is contaminated by the disease,
$P(D')=0.95$ the probability that the patient is not contaminated by the disease,
$P(S|D)=0.85$ the probability that a patient who has the disease shows the symptom and
$P(S|D')=0.25$ the probability that a patient who hasn't the disease shows the symptom
According to Bayes's theorem, the posterior probability that a patient who shows the
symptom is contaminated by the disease can be calculated as follows:
\begin{equation}\label{Bayes_th}
P(D|S)=\frac{P(S|D)P(D)}{P(S|D)P(D)+P(S|D')P(D')}\,. 
\end{equation}
Thus we have that $P(D|S)=\frac{0.85×0.05}{0.85×0.05+0.25×0.95}=0.15$,
which means that
there is only a 15\% chance that the patient examined has the disease even
if he presents a highly diagnostic symptom.

The inverse fallacy \cite{inverse fallacy}  is the erroneous assumption that $P(D|S)=P(S|D)$.
It is also called conversion error \cite{Wolfe}, the confusion hypothesis \cite{Macchi}, the Fisherian
algorithm \cite{Gigerenzer}. the conditional probability fallacy and the prosecutor's fallacy
\cite{prosecutor}. In particular, this last name evidences the fact (noted in many
studies) that both expert and non-expert 
judges often confuse a given
conditional probability with its inverse probability.

In a different filed, \cite{Meehl} evidences that clinicians consider
that the probability of the presence of a symptom given the diagnosis of a
disease is on its own a valid criterion for diagnosing the disease in the
presence of the symptom. This result has been later experimentally demonstrated
in  \cite{Hammerton, Liu},  where it has been found that the median judgments of
$P(D|S)$ is almost equal to the presented value of the
inverse probability, $P(S|D)$.
Similarly, \cite{Eddy} investigated how physicians estimated the probability that a
woman has breast cancer, given a positive result of a mammogram.
Approximately 95\% of clinicians surveyed gave a numerical answer close to
the inverse probability.

\section{Base-rate fallacy}\label{Base-rate fallacy}
The base rate fallacy \cite{base rate}, also called base rate neglect, is a logical fallacy that 
occurs when irrelevant information is used to make a probability judgment, 
especially when empirical statistics about the probability are available 
(called the \textit{base rate} or \textit{prior probability}).
In a study done by Tversky and Kahneman, subjects were given the following problem:
a cab was involved in a hit and run accident at night. Two cab companies, the Green and the Blue, 
operate in the city. 85\% of the cabs in the city are Green and 15\% are Blue.
A witness identified the cab as Blue. The court tested the reliability of the witness under the 
same circumstances that existed on the night of the accident and concluded that the witness correctly 
identified each one of the two colors 80\% of the time and failed 20\% of the time.
What is the probability that the cab involved in the accident was Blue rather than Green?

We define the following events:
$D$=Datum: there was an accident.
$H$=Hypothesis to evaluate: it was a blue cab.
From Bayes' theorem, the correct answer

\begin{eqnarray}
P(H|D)&=\frac{P(D|H)P(H)}{P(D|H)P(H)+P(D/H')P(H')}\\\nonumber
&=\frac{0,80 \cdot 0,15}{0,80 \cdot 0,15 + 0,20 \cdot 0,85}=0.41
\end{eqnarray}

Most subjects gave probabilities over 50\%, and some gave answers over 80\%. 
According to the base-rate fallacy, this can be explained with the fact that 
people tend to neglect the base rate, or prior probability, $P(H)$.

Some researchers have considered the inverse fallacy as the result of people's
tendency to consistently undervalue if not ignore the base-rate information
presented as a proxy for prior probabilities (e.g., \cite{Dawes, Kahn1973, Pollard}).
Other
researchers, however, have proposed that the base-rate effect was in fact
originating from the inverse fallacy and not the reverse (e.g., 
\cite{inverse fallacy, Wolfe}). In support of this notion, in \cite{Wolfe},
Experiment 3, it was found that participants who were trained to distinguish
$P(D|H)$ from $P(H|D)$ were less likely to exhibit base rate neglect compared
to a control group.  

\section{Quantum description of the inverse fallacy}
%
In this section we introduce the basic formalism of quantum mechanics: we will
not give an exhaustive description of the formalism, we will only give the necessary notions.
Our main hypothesis is that the estimated probabilities of agents relevant to a certain event
can be deduced from the quantum formalism. 
We now give some important definitions, valid both in classic and in quantum formalism:
\\
1) An \textit{observable} $A$ is an event which can be verified. For any observable $A$ we can 
always write the dichotomous question "is $A$ true?". In the following, we identify the event $A$ 
with the answer Yes, and we call $A'$ the answer No, or the negation of the event.
\\
2) The \textit{preparation} is any information previously given to the agent
which can be used to determine the estimated probabilities.
\\
3) the \textit{opinion state} (or simply state) of an agent is the result of the preparation.
\\
4) $P(A)$ is the \textit{estimated probability} that the event is true, given a set of agents
in a particular state. Analogously, we call
$P(A')$ the estimated probability that the event is false. Of course, $P(A)+P(A')=1$.

If the agents know that $A$ is true,
the resulting estimated probability is $P(A)=1$ and $P(A')=0$. In this case,
the preparation of the opinion state is the information "$A$ is true".

In the quantum case, the statistic description of the event $A$ is more complicated, and involves
the mathematical formalism of a complex separable Hilbert space $H$. The main concept is that
the opinion state of agents is described by a vector in a particular complex separable Hilbert space $H$.
For a dichotomous question, the dimension of the $H$ is 2, which gives the simplest quantum system,
the \textit{qubit} (containing the unit of quantum information). 

In order to describe a vector in the space $H$, we now give the standard bra-ket notation usually used 
in quantum mechanics,  introduced by Dirac \cite{Dirac}. According with this notation, a vector in $H$ is called a 
\textit{ket}, and written as $|s\rangle$. This vector describes the opinion state of agents.
A simple example of a ket can be given in the case that the agents already know that the observable $A$ is true/false: 
in the true case, the resulting vector is called $|A\rangle$, while in the false case $A'\rangle$.
Since $A$ excludes $A'$ (the observable can not be observed simulaneously true and false), 
this condition must be written in terms of a property of vector spaces: 
$|A\rangle$ and $|A'\rangle$ are orthogonal vectors, and thus form a complete basis of $H$.
This entails that any opinion state $|s\rangle$ can be written as a linear combination of
$\{|A\rangle, |A'\rangle\}$:
\begin{equation}\label{superposition_A} |s\rangle=s_0 |A\rangle + s_1 |A'\rangle\,,
\end{equation}
with $s_0$ and $s_1$ complex numbers. We also say that the  state
$|s\rangle$ is a \textit{superposition} of the basis states $\{|A\rangle, |A'\rangle\}$.

The orthogonality can be descrived mathematically with the concept of \textit{inner product}:
the inner product of two kets
$|s\rangle$ and $|s'\rangle$ can be written, in the basis $\{|A\rangle, |A'\rangle\}$, as 
$\langle s'|s\rangle=s_0 {s'_0}^*+s_1 {s'_1}^*$, where $s'_i$ are the
components of $|s'\rangle$ in the same basis. Thus the  inner
product of $|s\rangle$ and its dual vector $\langle s|$ is $\langle
s|s\rangle=|s_0|^2+|s_1|^2$, and it is equal to 1 if the vector is
normalized.
According to the bra-ket notation, the \textit{inner product} $\langle s'|s\rangle$  forms a "braket".
We now give three important properties of the inner product:
\\
a) for two any mutually exclusive events $A$, $A'$, $\langle A|A'\rangle=0$
\\
b) $\langle A|A\rangle=1$ (normalization property)
\\
c) given an opinion state $|s\rangle$ and an observable $|A\rangle$, the estimed probability of $A$ is
$P(A)=|\langle s|A\rangle|^2=|\langle A|s\rangle|^2$.

From these properties, we have that $\{|A\rangle, |A'\rangle\}$ form an ortonormal basis of $H$,
and they are useful in order to compute estimed probabilities.

The state $|s\rangle$ is called a \textit{pure} state, and
describes a quantum state for which the preparation is complete: all
the information which can be theorically provided have been used. Moreover,
from the properties of the inner product we can write
\begin{equation}\label{superposition_A_estimed} 
|s\rangle=\sqrt{P(A)} |A\rangle + \sqrt{A'}e^{i\phi_a}|A'\rangle\,,
\end{equation}
where $P(A),P(A')$ are the estimed probability, and $e^{i\phi_a}$ is a phase, important in interference effects (for example, see the conjunction fallacy \cite{Rrfranco2}).

A mixed state describes a situation where there are different complete preparations $|s_i\rangle$, but each associated with a probability $P_i$. This means that we can find some agents within an opinion state $|s_i\rangle$ with a probability $P_i$. The mixed states are not represented as vetors, but as matrices, the density matrix $\rho$:
\begin{equation}\label{mixed} 
\rho=\sum_i P_i|s_i\rangle\ \langle s_i|,,
\end{equation}
And important property of $\rho$ is that its diagonal elements in the basis of $A$ are exactly $P(A)$ and $P(A')$.
\subsection{Observables and collapse}
Let us suppose that the preparation of the opinion state $P(A)$ of the agents is such that $P(A)<1$.
After the preparation, if they know that $A$ is true, the resulting estimated probability becomes 
$P(A)=1$ and $P(A')=0$. The knowledge of agents about $A$ has increased.

In the quantum formalism we have a similar fact: given a 
preparation of the opinion state $|s\rangle$ of the agents such that $|\langle s|A\rangle|^2<1$, if the agents 
know that $A$ is true, the resulting estimated probability becomes $P(A)=1$, and the 
resulting state is $|A\rangle$. This is called the \textit{collapse} of the quantum state into a base vector,
and has many important consequances in the interpretations of quantum theory.
%
\subsection{Conditional probability and inverse fallacy}
We now consider two questions $A$ and $B$, with the corresponding basis states
$\{|A\rangle, |A'\rangle\}$ and $\{|B\rangle, |B'\rangle\}$. Moreover, we suppose that $0 < |\langle A|B\rangle|^2 < 1$, which means that these vectors are not parallel (and thus do not represent the same question).

We can compute the conditional probability $P(A|B)$ by noting that the previous knowledge of the answer $B$ entails that the actual state is $|B\rangle$ (see the collapse of state vector).
Thus we have that 
\begin{equation}\label{conditional}
P(A|B)=|\langle A|B\rangle|^2=|\langle B|A\rangle|^2=P(B|A)\,,
\end{equation}
which evidences that the quantum formalism naturally leads to the inverse fallacy.
This equality is always valid in quantum formalism, both for pure and mixed states.
The fact that the inverse fallacy manifests itself only in certain situations can 
be explained with the presence of a bounded-rationality regime (which is the analogue of the quantum regime). 

In the present article we are not interested in the situations which lead to the presence of such a regime; in fact, differently from other cognitive heuristics, the 
inverse fallacy is always entailed by the quantum formalism: there are no states without inverse fallacy.

\subsection{Additivity}
First of all we note that the completeness of the basis of $H$ entails the 
simple condition
\begin{equation}\label{additivity}
P(A|B)+P(A'|B)=1\,.
\end{equation}
Equation (\ref{conditional}) of the quantum formalism entails the following 
equality, which is identical to the additivity principle: 
\begin{equation}\label{additivity1}
P(B|A)+P(B|A')=1\,.
\end{equation}

It is important to note that equation (\ref{additivity1}) concerns the estimated probabilities, while for real probabilities we have
\begin{equation}\label{non-additivity}
P(B|A)+P(B|A')=\frac{P(A|B)P(B)}{P(A)}  +   \frac{P(A'|B)P(B)}{P(A')}\,,
\end{equation}
which can be higher or lower than 1 (see \cite{inverse fallacy}).
There are  experimental situations \cite{inverse fallacy} where agents judge  $P(A|B)=P(B|A)$, i.e. they commit the inverse fallacy, but the given data are the probabilties $P(B|A)$ and $P(B|A')$, whose sum can be higher or lower than 1. Thus the agents commit inverse fallacy and judge $P(A|B)=P(B|A)$,  violating equation (\ref{additivity}). The problem is that the initial probabilities $P(B|A)$ and $P(B|A')$ have been provided as initial data, and have not been estimated by the agents.
In this case we say that the given information are not consistent with the quantum regime.

Finally the quantum formalism, through equation (\ref{additivity1}),
leads to the following equalities
\begin{equation}\label{additivity2}
P(A|B')=P(A'|B)\,\,,\,\, P(A|B)=P(A'|B')\,.
\end{equation}
This can be easily shown, for example in the first case, by noting that
$P(B'|A)=P(A|B')=1-P(A|B)=1-P(B|A)=P(B'|A)$.

Thus we have found very simple equations that link the conditional probabilities,
valid only in bounded-rationality regime: this confirms our hypothesis, that 
agents in such a regime tend to estimate probabilities in an intuitive way.

\subsection{Change of basis}
The additivity conditions (\ref{additivity}), (\ref{additivity1}) and  (\ref{additivity2}) are consistent with the quantum formalism. This can be seen also by considering the unitary operators which represents the change of basis from  $\{|A\rangle, |A'\rangle\}$ to  $\{|B\rangle, |B'\rangle\}$ and viceversa:
\begin{eqnarray} \label{B-A-basis}
|B'\rangle=e^{-i\xi}\sqrt{P(B|A)}|A'\rangle-e^{i\phi}\sqrt{P(B|A')}|A\rangle\\\nonumber
|B\rangle=e^{-i\phi}\sqrt{P(B|A')}|A'\rangle+e^{i\xi}\sqrt{P(B|A)}|A\rangle\,\,.
\end{eqnarray}
The transformations above are a change of basis, which can be described in
terms  of a special unitary operator $\widehat{U}$ (which preserves the
normalization) such that $\sum_{ij}U_{ij}|a_i\rangle=|b_j\rangle$. The
transformation is defined uniquely by the three independent parameters
$P(B|A)$, $\phi$ and $\xi$: in fact the other parameter
$P(B|A')=1-P(B|A)$ is not independent. However, this trasformation
is consistent with the definition of conditional probability (\ref{conditional}), as can be easily seen by using equation (\ref{additivity2}).

\section{conclusions}\label{conclusions}
In this article we have presented a description in the quantum formalism 
of the inverse fallacy: we have shown that this fallacy is a natural and 
very general consequence of the quantum formalism and of the Hilbert space 
laws.

This result makes stronger the point of view of quantum cognition, which
studies the behavour of agents in bounded-rationality regime. In particular, 
it is important to note that the predictions of the quantum cognition
are relevant to the estimated probabilities, which follow different laws from
the classic probability theory.
 \footnotesize
%
%

\end{document}